# A novel protocol for linearization of the Poisson-Boltzmann equation

Roumen Tsekov
Department of Physical Chemistry, University of Sofia, 1164 Sofia, Bulgaria

A new protocol for linearization of the Poisson-Boltzmann equation is proposed and the resultant electrostatic equation coincides formally with the Debye-Hückel equation, the solution of which is well known for many electrostatic problems. The protocol is examined on the example of electrostatically stabilized nano-bubbles and it is shown that stable nano-bubbles could be present in aqueous solutions of anionic surfactants near the critical temperature, if the surface potential is constant. At constant surface charge non nano-bubble could exist.

The theory of the electric double layer [1] dates back to the classical works of Helmholtz, Gouy, Chapman, Stern, Debye and Hückel. Due to the theoretical complications in dense systems [2] the application is mainly restricted to dilute ionic solutions, where the electrostatic potential is described via the Poisson-Boltzmann equation. The interaction between electric double layers has been also studied intensively as an important component of the particle and colloidal forces [3]. Significant attention has been paid recently to highly-charged Coulomb mixtures, where many specific phenomena take place [4] among them mono-species electric double layers [5]. General theories of charged fluids are developed [6-9], which account for ion correlations going beyond the Poisson-Boltzmann theory. An interesting effect here is the non-electrostatic interaction between the ions in an electric double layer expected to become important in concentrated solutions [2]. The aim of the present paper is to develop a new protocol for linearization of the Poisson-Boltzmann equation, which could be useful for solving easily electrostatic problems in colloidal systems. As an example, the protocol is applied for description of electrostatically stabilized nano-bubbles [10].

In electrostatics of disperse systems one needs regularly to solve the Poisson equation [1]

$$\varepsilon_0 \varepsilon \Delta \phi = -\rho \qquad (1)$$

governing the electrostatic potential $\phi$. The charge density in the solution $\rho(\phi)$ is modelled often via the Boltzmann distribution. Due to the mathematical complexity, the nonlinear Poisson-Boltzmann equation is solvable numerically only. However, Eq. (1) is analytically solved in practice for many colloidal systems in the frames of the Debye-Hückel linear approximation

$$\Delta\phi = \kappa_D^2 \phi \tag{2}$$

where $\kappa_D = \sqrt{-(\partial_\phi \rho)_{\phi=0}/\varepsilon_0\varepsilon}$ is the reciprocal Debye length. This equation is derived via a standard linearization of Eq. (1) for low potentials. In the present paper a new protocol for linearization is proposed, which seems to be more general and useful for arbitrary potentials.

Introducing a constant electrostatic potential $\phi_0$, being typical for the considered disperse system, one can approximate Eq. (1) as follows

$$\Delta\phi = \kappa^2 \phi \tag{3}$$

where $\kappa \equiv \sqrt{-\rho(\phi_0)/\varepsilon_0\varepsilon\phi_0}$ is a new screening parameter. In fact, Eq. (3) could be applied to an arbitrary charge density distribution and the Boltzmann distribution is simply an example here. Note that $\kappa(\phi_0 = 0) = \kappa_D$ and thus the Debye-Hückel approximation is a particular case of Eq. (3). The advantage of this new protocol as compared to the standard linearization scheme $\rho(\phi) \approx \rho(\phi_0) + (\partial_\phi \rho)_{\phi=\phi_0}(\phi - \phi_0)$ is that Eq. (3) coincides formally with Eq. (2), which is solved elsewhere. Thus, the solution of Eq. (3) is well known and, for instance, it reads to

$$\phi = \phi_s \exp[\kappa(R-r)](R/r) \quad \phi = \phi_s \exp(-\kappa x) \quad \phi = \phi_s \cosh(\kappa z)/\cosh(\kappa h/2) \tag{4}$$

near a spherical particle with radius $R$, a flat surface and in a symmetric film with thickness $h$, respectively. These solutions certainly differ from the exact solutions of the nonlinear Eq. (1) but varying the parameter $\phi_0$ one can always find a good approximation. For example, if one is looking primarily for the correct surface potential $\phi_s$ it is reasonable to accept $\phi_0 = \phi_s$, while in the case of interest on the overall dependence of the electric potential the choice $\phi_0 = \phi_s/2$ seems more plausible. Finally, a mean-field approximation is also possible, which consists in replacement of $\phi_0$ by $\phi$ in the final solutions (4). Thus, for example, the mean-field potential on a flat surface of 1:1 electrolyte will be the solution of the following transcendental equation

$$\phi = \phi_s \exp[-\sqrt{(k_B T/e\phi)\sinh(e\phi/k_B T)}\,\kappa_D x]$$

For this particular case the exact solution of the nonlinear Poisson-Boltzmann equation is known, $\kappa_D x = 2\operatorname{arctanh}[\exp(e\phi/2k_B T)] - 2\operatorname{arctanh}[\exp(e\phi_s/2k_B T)]$. It is plotted in Fig. 1 together with the mean-field approximation in the form $\kappa_D x = -\ln(\phi/\phi_s)/\sqrt{(k_B T/e\phi)\sinh(e\phi/k_B T)}$ and the Debye-Hückel approximate solution $\kappa_D x = -\ln(\phi/\phi_s)$. As is seen, the mean-field solu-tion almost follows the exact solution near the surface, while at long distance it tends asymptotically to the Debye-Hückel approximation. Thus, the juxtaposition is very good at larger potentials, which are important for the specific electric properties on the boundary interfaces. Hence, the present new protocol for linearization of the Poisson-Boltzmann equation seems to be quite good and captures the essential physics of the charged system.

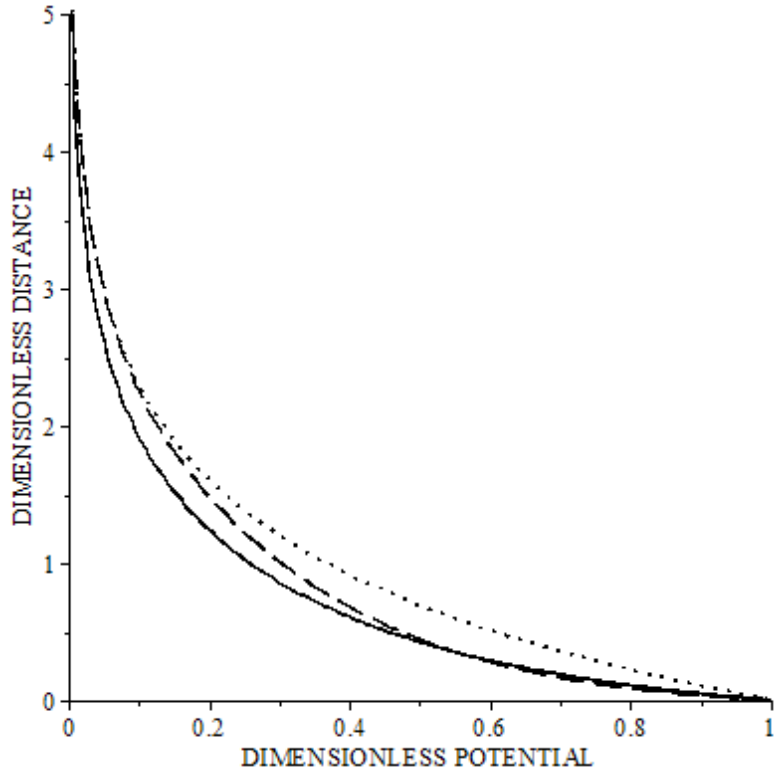

**Fig. 1** The dependence of the dimensionless distance $\kappa_D x$ on the dimensionless potential $\phi/\phi_s$ for 1:1 electrolyte at a high surface potential $\phi_s = 5k_B T/e$: the exact solution (solid line), the mean-field approximation (dashed line) and the Debye-Hückel approximation (dotted line)

In order to get advantage of the new protocol for linearization of the Poisson-Boltzmann equation, let us consider a nano-bubble with a radius $R$ immersed in aqueous electrolyte solution. It is assumed that ions can adsorb at the bubble interface. Since there are no ions in the gas phase, the electrostatic potential inside the nano-bubble is constant. In the liquid $\phi$ satisfies the Poisson-Boltzmann equation (1), which, written in spherical coordinates, reads

$$\varepsilon_0 \varepsilon \partial_r (r^2 \partial_r \phi)/r^2 = -\rho(\phi) \tag{5}$$

As was mentioned, Eq. (5) can be analytically solved for low potentials but in the present case we are looking for strong electrostatic effects due to huge surface potentials. Thus, the Debye-Hückel approximation is not applicable. Fortunately, in this case the new protocol, valid for electric double layers being very condensed near the surface, is reasonable. Hence, close to the surface the electrostatic potential is of the order of $\phi_s$ and the Poisson-Boltzmann equation (5) can be linearized in the form

$$\partial_r(r^2 \partial_r \phi)/r^2 = \kappa_s^2 \phi \tag{6}$$

where $\kappa_s \equiv \sqrt{-\rho(\phi_s)/\varepsilon_0 \varepsilon \phi_s}$. This equation is exact at the bubble surface. In the bulk Eq. (6) is not precise but the defect is not essential since the potential inside the solution is very low as compared to the surface potential $\phi_s$ due to strong electrostatic screening. As was mentioned, the advantage of Eq. (6) is that its analytical solution is known

$$\phi = \phi_s \exp[\kappa_s (R-r)](R/r) \tag{7}$$

The surface potential $\phi_s$ on the bubble/water interface is related to the surface charge density $q_s$ via the relation [1]

$$q_s \equiv -\varepsilon_0 \varepsilon (\partial_r \phi)_{r=R} = \varepsilon_0 \varepsilon \phi_s (\kappa_s + 1/R) \tag{8}$$

where the last expression is obtained by employing Eq. (7). The presence of the electrostatic potential in the liquid generates additional pressure on the bubble via the Maxwell stress tensor [1]. Hence, the normal and tangential components of the pressure tensor in the liquid read

$$P_N = p_L + \varepsilon_0 \varepsilon \kappa_s^2 \phi^2 /2 - \varepsilon_0 \varepsilon (\partial_r \phi)^2 /2 \qquad P_T = p_L + \varepsilon_0 \varepsilon \kappa_s^2 \phi^2 /2 + \varepsilon_0 \varepsilon (\partial_r \phi)^2 /2 \tag{9}$$

respectively, where $p_L$ is the constant molecular pressure far away from the bubble. The second term here represents the osmotic pressure due to the differences in the ionic concentrations. One can easily check that these pressure components satisfy the equilibrium condition $\partial_r(r^2 P_N) = 2r P_T$. Integrating the latter yields the excess electrostatic force acting on the nano-bubble

$$\pi R^2 (P_{N,s} - p_L) = \int_R^\infty (p_L - P_T) 2\pi r dr = -\pi \varepsilon_0 \varepsilon \phi_s^2 (\kappa_s R + 1/2) \tag{10}$$

where $P_{N,s} = P_N(R)$ is the pressure on the bubble surface. As is seen, this excess force is negative, thus leading to stabilization of the nano-bubble. On the other hand, the normal pressure balance on the bubble surface reads

$$p_G - P_{N,s} = 2\sigma_0 / R \tag{11}$$

where $p_G$ is the gas pressure inside the bubble and $\sigma_0$ is the constant surface tension on a flat liquid/gas interface at zero surface charge. Note that the electrostatic effects on the surface tension are already accounted for in Eq. (10).

Due to the surrounding atmosphere, the aqueous solution is in equilibrium with a bulk gas phase as well. Therefore, $p_G = p_L$ and Eq. (11) accomplished by Eq. (10) leads to an expression for the equilibrium radius of the bubble

$$R = \varepsilon_0 \varepsilon \phi_s^2 / 4(\sigma_0 - \varepsilon_0 \varepsilon \kappa_s \phi_s^2 / 2) \tag{12}$$

In the colloidal science there are two well-known interfacial models [11]: the constant surface charge density and the constant surface potential. In the first case $q_s = q_\infty$ and, hence, according to Eq. (8) the surface potential equals to $\phi_s = q_\infty / \varepsilon_0 \varepsilon (\kappa_s + 1/R)$. Since we are looking for nano-bubbles with very small radii it follows from this relation that the surface potential is also very low. Hence, $\kappa_s \approx \kappa_D$ and Eq. (12) can be rewritten in the form

$$(1 + 1/\kappa_D R)^2 / (1 + 1/2\kappa_D R) = q_\infty^2 / 2\varepsilon_0 \varepsilon \kappa_D \sigma_0 \tag{13}$$

This equation has a positive solution for $R$ only if $q_\infty^2 \geq 2\varepsilon_0 \varepsilon \kappa_D \sigma_0$. However, since the surface tension on a flat liquid/gas surface equals to $\sigma_\infty = \sigma_0 - q_\infty^2 / 2\varepsilon_0 \varepsilon \kappa_D > 0$, the inequality above is never satisfied. Therefore, the conclusion is that no stable nano-bubble is possible at constant surface charge density.

On the contrary, at constant surface potential $\phi_s = \phi_\infty$ Eq. (12) reads

$$R = \varepsilon_0 \varepsilon \phi_\infty^2 / 4(\sigma_0 - \varepsilon_0 \varepsilon \kappa_s \phi_\infty^2 / 2) = \varepsilon_0 \varepsilon \phi_\infty^2 / 4\sigma_\infty \tag{14}$$

Hence, the nano-bubble radius is larger as higher the surface potential and lower the surface tension are. A standard way to try to achieve such conditions is addition ionic surfactants. However, traditional ionic surfactants cannot lower the surface tension enough to stabilize the nano-bubbles. If we accept the zeta-potential of pure water $\zeta = -65$ mV [12] as the value of surface potential, a nano-bubble with radius $R=1$ nm would correspond to $\sigma_\infty = 1$ mN/m according to Eq. (14). Obviously, to reach this very low value of surface tension one has to play with temperature. Therefore, stable nano-bubbles with radii of several nanometers are expected to be present in water near the critical temperature, especially at presence of anionic surfactants.